\documentclass[aps,showpacs,prb]{revtex4}
\usepackage{graphicx}
\usepackage{amsmath}

\begin{document} 

\title{Many-body theory of electronic transport in single-molecule heterojunctions}

\author{J.~P.~Bergfield}
\affiliation{College of Optical Sciences, University of Arizona, 1630 East University Boulevard, AZ 85721}
\author{C.~A.~Stafford}
\affiliation{Department of Physics, University of Arizona, 1118 East Fourth Street, Tucson, AZ 85721}


\date{\today}

\begin{abstract}
A many-body theory of molecular junction transport based on nonequilibrium Green's functions is developed, which treats coherent quantum effects
and Coulomb interactions on an equal footing.  The central quantity of the many-body theory is the Coulomb self-energy matrix
$\Sigma_{\rm C}$ of the junction.  $\Sigma_{\rm C}$ is evaluated exactly in the sequential tunneling limit, and the correction due to 
finite tunneling width is evaluated self-consistently using a conserving approximation based on diagrammatic perturbation theory on the
Keldysh contour.
Our approach reproduces the key features of both the Coulomb blockade and coherent 
transport regimes simultaneously in a single unified transport theory.
As a first application of our theory, we have calculated the thermoelectric power and differential conductance spectrum
of a benzenedithiol-gold junction using a semi-empirical $\pi$-electron Hamiltonian that accurately describes the
full spectrum of electronic excitations of the molecule up to 8--10eV.
\end{abstract}

\maketitle

\section{Introduction}

Electron transport in single-molecule junctions 
\cite{Nitzan03,Natelson06,Tao06} 
is of fundamental interest as 
a paradigm for {\em nanosystems far from equilibrium},
and as a means to probe 
important chemical\cite{BioBook99} 
and biological\cite{Schrieber05} processes, with myriad potential device applications.\cite{Tao06,Cardamone06,Stafford07b}
A general theoretical framework
to treat the 
many-body problem of 
a single molecule coupled to metallic electrodes does not currently exist.
Mean-field approaches based on density-functional theory
\cite{Todorov00,Diventra01,Guo01,Heurich02,Emberly03,Tomfohr04,Lindsay07}---the dominant paradigm in quantum chemistry---have serious shortcomings
\cite{Datta06,Ke07,Cohen08}
because they do not account for important interaction effects like 
Coulomb blockade.

An alternative approach is to solve the few-body molecular Hamiltonian exactly, and treat electron hopping
between molecule and 
electrodes as a perturbation.
This approach has been used 
to describe molecular junction
transport in the sequential-tunneling regime,\cite{Schoeller03,Datta06,Begemann08} but
describing coherent quantum transport in this framework remains an open theoretical problem.
Higher-order tunneling processes may be treated rigorously
in the density-matrix formalism,\cite{Koenig97,Schoeller00}
but the expansion is typically truncated 
at fourth order,\cite{Koenig97,Wacker05} the calculation of higher-order terms being prohibitively difficult. 

In this article, we
develop a many-body theory of molecular junction transport based on
nonequilibrium Green's functions\cite{Meir92,Jauho96,Viljas05} (NEGF), 
in order to utilize physically motivated approximations that sum terms of all orders.
The junction Green's functions are calculated exactly in the sequential-tunneling limit, 
and the corrections to the electron self-energy due to finite
tunneling width 
are included via Dyson-Keldysh equations.  
The {\em tunneling self-energy} is calculated exactly 
using the equations-of-motions method,\cite{Meir91,Jauho96} while the correction to the {\em Coulomb self-energy} is calculated 
using diagrammatic perturbation theory.
In this way, tunneling processes 
are included to infinite order, meaning that any approximation utilized
is a truncation in the physical processes considered rather than in the order of those processes.

Our approach reproduces the key features of both the Coulomb blockade and coherent 
transport regimes simultaneously in a single unified transport theory.
Nonperturbative effects of intramolecular correlations are included,
which are necessary to accurately describe the HOMO-LUMO gap, essential for a quantitative theory of transport. 

As a first application of our many-body transport theory, we investigate 
the benchmark system of benzene(1,4)dithiol (BDT) with gold electrodes.
Two key 
parameters determining the lead-molecule coupling---the tunneling width $\Gamma$ and the chemical potential offset $\Delta \mu$---are
fixed by comparison to linear-response measurements of the thermoelectric power\cite{Reddy07,Baheti08} and electrical 
conductance.\cite{Xiao04}
The nonlinear junction response is then calculated.
The differential conductance spectrum of the junction exhibits 
an irregular ``molecular diamond'' structure analogous to the
regular Coulomb diamonds\cite{DeFranceschi01} observed in quantum dot transport experiments, 
as well as clear signatures of 
coherent quantum transport---such as transmission nodes 
due to destructive interference---and 
of resonant tunneling through molecular excited states.

The article is organized as follows:  A detailed derivation of the 
many-body theory of molecular junction transport is presented in Sec.\ \ref{sec:manybody_formalism}. 
Two useful approximate solutions for the nonequilibrium Coulomb self-energy are given in Secs.\ \ref{sec:cotunneling} and \ref{sec:HF}.
The details of the model used to describe 
a molecular heterojunction consisting of a $\pi$-conjugated molecule covalently bonded to 
metallic electrodes are presented in Sec.\ \ref{sec:molecular_model}.
The electric and thermoelectric response of a BDT-Au junction are calculated 
in Sec.\ \ref{sec:BDT}, and compared to experimental results.
A discussion and conclusions are presented in Sec.\ \ref{sec:conclusion}.

\section{Nonequilibrium many-body formalism}
\label{sec:manybody_formalism}

\subsection{Molecular junction Hamiltonian}
\label{sec:H_general}

The Hamiltonian of a junction consisting of a molecule coupled to several 
metallic electrodes may be written
\begin{equation}
H_{\rm junction}=H_{\rm mol} + H_{\rm leads} + H_{\rm T}.
\label{eq:H_junction}
\end{equation}
The molecular Hamiltonian can be formally divided into one-body and two-body terms $H_{\rm mol}=H^{(1)}_{\rm mol}+ H^{(2)}_{\rm mol}$.
In general, neglecting spin-orbit coupling, the one-body term can be written
\begin{equation}
H^{(1)}_{\rm mol} = \sum_{n,m,\sigma} \left[H^{(1)}_{\rm mol}\right]_{n\sigma,m\sigma} d_{n\sigma}^\dagger d_{m\sigma},
\label{eq:H1_form}
\end{equation}
where $d^\dagger_{n\sigma}$ creates an electron of spin $\sigma$ on atomic orbital $n$ of the molecule
and $[H^{(1)}_{\rm mol}]$ is a hermitian matrix.
For simplicity, the atomic basis orbitals are taken to be orthonormal, so that the anticommutator
$\{d_{n\sigma}^\dagger, d_{m\sigma'}\}=\delta_{nm}\delta_{\sigma\sigma'}$.
Extension to non-orthogonal bases is straightforward in principle.
For a general spin-rotation invariant two-body (e.g. Coulomb) interaction, 
\begin{equation}
H^{(2)}_{\rm mol} = \frac{1}{2}\sum_{n,m} U_{nm} \rho_n \rho_m,
\label{eq:H2_form}
\end{equation}
where $\rho_n=\sum_\sigma d_{ n\sigma}^\dagger d_{n\sigma}$.

Due to their large density of states (and consequently good screening), the macroscopic metallic electrodes 
(labeled $\alpha\in[1,\ldots,M]$) may be modeled as non-interacting Fermi gases: 
\begin{equation}
H_{\rm leads} = \sum_{\alpha=1}^M\sum_{\substack{k\in\alpha\\ \sigma}}\epsilon_{k\sigma} c_{k\sigma}^\dagger c_{k\sigma},
\label{eq:Hleads}
\end{equation}
where $c^\dagger_{k\sigma}$ creates an electron of energy $\epsilon_{k\sigma}$ in lead $\alpha$.
The electrostatic interaction 
of molecule and electrodes due to the electric dipoles formed at each molecule-electrode interface may be included 
in $H^{(1)}_{\rm mol}$, as discussed in Sec.\ \ref{sec:molecular_model}.
Tunneling of electrons between the molecule and the electrodes is described by the Hamiltonian
\begin{equation}
H_{\rm T} =\sum_{\alpha=1}^M \sum_{k\in\alpha} \sum_{n,\sigma} \left(V_{nk}d_{ n\sigma}^\dagger c_{k\sigma}+\mathrm{H.c.}\right). 
\label{eq:Htun}
\end{equation}

\subsection{Non-equilibrium Green's functions} 
\label{sec:NEGF}

The electronic system (\ref{eq:H_junction}) of molecule 
plus electrodes 
has an infinite Hilbert space. 
A formal simplification 
of the problem is obtained through the use of the Green's functions\cite{Meir92,Jauho96} 
\begin{eqnarray}
G_{n\sigma,m\sigma'}(t) & = & -i\theta(t)\langle \{d_{n\sigma}(t),d_{m\sigma'}^\dagger(0)\}\rangle,\\
G^<_{n\sigma,m\sigma'}(t) & = & i\langle d_{m\sigma'}^\dagger(0)\, d_{n\sigma}(t) \rangle,
\end{eqnarray}
known as the retarded and Keldysh ``lesser'' Green's functions, respectively.
Physical observables in the molecular transport junction can be expressed 
in terms of $G$ and $G^<$.
In this section, an exact Dyson equation for $G$ 
is derived, along with a corresponding Keldysh equation for 
$G^<$.  

Setting $\hbar=1$,
the retarded Green's function obeys the equation of motion \cite{Jauho96} 
\begin{equation}
i\frac{\partial}{\partial t} G_{n\sigma,m\sigma'}(t) = \delta(t)\delta_{nm}\delta_{\sigma\sigma'}
- i\theta(t)\left\langle\left\{\left[d_{n\sigma}(t),H_{\rm junction}\right],d^\dagger_{m\sigma'}(0)\right\}\right\rangle.
\label{eq:Gr_EOM}
\end{equation}
For the purposes of this article, only the spin-diagonal term $G_{n\sigma,m\sigma}$ is needed.
Evaluating the commutators $[d_{n\sigma},H_{\rm mol}]$ and $[d_{n\sigma},H_{\rm T}]$, and noting that $[d_{n\sigma},H_{\rm leads}]=0$, 
Eq.\ (\ref{eq:Gr_EOM}) becomes
\begin{equation}
i\frac{\partial}{\partial t} G_{n\sigma,m\sigma}(t) = \delta(t)\delta_{nm}
+\sum_{n'} \left[H^{(1)}_{\rm mol}\right]_{n\sigma,n'\sigma} G_{n'\sigma,m\sigma}(t)
+ \sum_\alpha \sum_{k\in\alpha} V_{nk} \, g_{k\sigma, m\sigma}(t)
+ \sum_{n'} U_{nn'} G^{(2)}_{n',n\sigma m\sigma}(t),
\label{eq:Gr_EOM2}
\end{equation}
where 
\begin{equation}
	g_{k\sigma, m\sigma'}(t) = -i\theta(t)\left\langle\left\{c_{k\sigma}(t),d^\dagger_{m\sigma'}(0)\right\}\right\rangle
	\label{eq:Lead_Mol_GreenFcn}
\end{equation}
and
\begin{equation}
G^{(2)}_{n',n\sigma m\sigma'}(t) = -i\theta(t)\left\langle\left\{\rho_{n'}(t)d_{n\sigma}(t),d^\dagger_{m\sigma'}(0) \right\}\right\rangle.
\label{eq:G2}
\end{equation} 
The equation of motion for $g_{k\sigma, m\sigma}(t)$ is 
\begin{equation}
i\frac{\partial}{\partial t} g_{k\sigma, m\sigma}(t) = 
- i\theta(t)\left\langle\left\{\left[c_{k\sigma}(t),H_{\rm junction}\right],d^\dagger_{m\sigma}(0)\right\}\right\rangle.
\label{eq:g_EOM}
\end{equation}
Evaluating the commutators $[c_{k\sigma},H_{\rm T}]$ and $[c_{k\sigma},H_{\rm leads}]$, and noting that $[c_{k\sigma},H_{\rm mol}]=0$,
Eq.\ (\ref{eq:g_EOM}) becomes 
 \begin{equation}
	i\frac{\partial}{\partial t} g_{k\sigma, m\sigma}(t) = \epsilon_{k\sigma} g_{k\sigma, m\sigma}(t)
                                                               +\sum_{n'} V^*_{n'k} G_{\rm n'\sigma,m\sigma}(t).
	\label{eq:g_EOM2}
\end{equation}

Fourier transforming Eqs.\ (\ref{eq:Gr_EOM2}) and (\ref{eq:g_EOM2}) into the energy domain, and eliminating $g_{k\sigma, m\sigma}(E)$, one arrives
at the following matrix equation for $G(E)$:
\begin{equation}
	\left[{\bf 1}E - H_{\rm mol}^{(1)} - \Sigma_{\rm T}\left(E\right) \right] G(E) = {\bf 1} + U G^{(2)}(E),
	\label{eq:NearlyFinal_EOM}
\end{equation} 
where the retarded {\em tunneling self-energy} matrix is
 \begin{equation}
 \left[\Sigma_{\rm T}(E)\right]_{n\sigma,m\sigma'}=\delta_{\sigma\sigma'}\sum_{\alpha}\sum_{ k \in \alpha} 
 \frac{V_{nk} V_{mk}^*}{E - \epsilon_{k\sigma} +i0^+}.
 \label{eq:SigmaRet_t}
 \end{equation}
Eq.\ (\ref{eq:NearlyFinal_EOM}) may be recast in the form of Dyson's equation
\begin{equation}
G(E) = \left[{\bf 1}E - H_{\rm mol}^{(1)} - \Sigma_{\rm T}(E) - \Sigma_{\rm C}(E)\right]^{-1}
\label{eq:Dyson}
\end{equation}
via the ansatz $UG^{(2)}(E)\equiv \Sigma_{\rm C}(E)G(E)$, which defines \cite{Mahan} the retarded {\em Coulomb self-energy} matrix $\Sigma_{\rm C}(E)$,
the central quantity of the many-body theory,
which must be determined via an appropriate approximation, as discussed below.

For nonequilibrium problems, the Keldysh ``lesser'' self-energy and Green's function are also needed.
$G^<$ is determined by the Keldysh equation \cite{Jauho96}
\begin{equation}
G^<(E)=G(E)\Sigma^<(E)G^\dagger(E),
\label{eq:Keldysh}
\end{equation}
where $\Sigma^<(E)=\Sigma_{\rm T}^<(E)+\Sigma_{\rm C}^<(E)$, and
$G^\dagger(E)$ is the Fourier transform of the advanced Green's function.
The ``lesser'' tunneling self-energy is 
\begin{equation} 
\left[\Sigma_{\rm T}^<(E)\right]_{n\sigma,m\sigma'}= i 
\sum_{\alpha}f_\alpha(E)\left[\Gamma^\alpha(E)\right]_{n\sigma,m\sigma'}^*,
\label{eq:SigmaLess_t}
\end{equation}
where 
$f_\alpha(E)=\{1+\exp[(E-\mu_\alpha)/k_B T_\alpha]\}^{-1}$ is the Fermi-Dirac distribution for lead $\alpha$, and
\begin{equation}
\left[\Gamma^{\alpha}(E)\right]_{n\sigma,m\sigma'}=2\pi\delta_{\sigma\sigma'} \sum_{k\in\alpha}V_{nk}^* V_{mk}
      \, \delta(E-\epsilon_{k\sigma})
\label{eq:GammaMatrix}
\end{equation}
is the tunneling-width matrix for lead $\alpha$.

Once the Green's functions are known, the relevant physical observables can be calculated.
For example, the current flowing into lead $\alpha$ is given by \cite{Meir92} 
\begin{equation}
I_\alpha = \frac{ie}{h} \int dE \; {\rm Tr}\left\{ \Gamma^\alpha(E)\left( G^<(E) + f_\alpha(E)\left[G(E)-G^\dagger(E)\right]\right) \right\}.
\label{eq:Meir_Wingreen}
\end{equation}

\subsection{Sequential tunneling limit} 
\label{sec:seq_tun}

\begin{figure}[tb]
\centering
\includegraphics[width=3.5in]{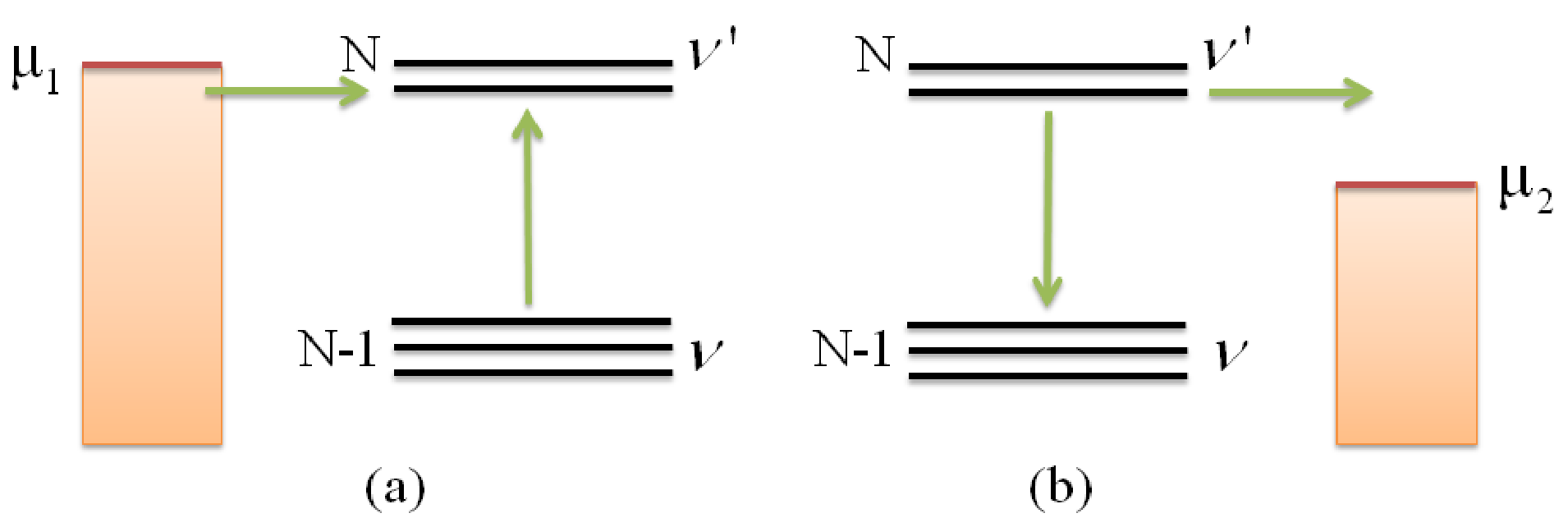} 
\caption[detailed balance]{A schematic representation of detailed balance.  The left-hand and right-hand sides of Eq.~(\ref{eq:detailed_balance}) 
are represented in (a) and (b), respectively.}
\label{fig:Detailed_balance}
\end{figure}

In the limit of infinitessimal lead-molecule coupling $\Sigma_{\rm T}/k_B T \rightarrow 0$,
coherent superpositions of different energy eigenstates of the molecule can be neglected,
and the junction Green's function becomes 
\begin{equation}
\lim_{\Sigma_{\rm T}\rightarrow 0} \left[G(E)\right]_{n\sigma,m\sigma} \equiv
\left[G_{\rm mol}(E)\right]_{n\sigma,m\sigma} = 
\sum_{\nu, \nu'} \left[{\cal P}(\nu) + {\cal P}(\nu')\right]
\frac{\langle \nu | d_{n\sigma} |\nu'\rangle \langle\nu'| d_{m\sigma}^\dagger |\nu\rangle}{E - E_{\nu'} + E_\nu + i 0^+},
\label{eq:Gintra}
\end{equation}
where $|\nu\rangle$ and $|\nu'\rangle$ are many-body eigenstates of the isolated molecule satisfying $H_{\rm mol} |\nu\rangle = E_\nu |\nu\rangle$, 
etc.  The nonequilibrium
probabilities ${\cal P}(\nu)$ can be determined by solving a system of semiclassical rate equations for sequential 
tunneling.\cite{Beenakker91,Schoeller03,Datta06,Begemann08} For
steady-state transport, they 
satisfy the principle of detailed balance (see Fig.\ \ref{fig:Detailed_balance}): 
\begin{equation}
{\cal P}(\nu) \sum_\alpha \tilde{\Gamma}^{\nu  \nu'}_{\alpha}\! f_\alpha \left( E_{\nu'}-E_\nu \right)
=
{\cal P}(\nu') \sum_\alpha \tilde{\Gamma}^{\nu  \nu'}_{\alpha}\! \left[1-f_\alpha \left( E_{\nu'}-E_\nu \right) \right].
\label{eq:detailed_balance}
\end{equation}
Here the rate constants are
given by Fermi's Golden rule as
\begin{equation}
\tilde{\Gamma}^{\nu  \nu'}_{\alpha} = {\rm Tr}\left\{ \Gamma^\alpha(E_{\nu'}-E_{\nu}) C^{\nu  \nu'}\right\},
\label{eq:FGR_detailed_balance}
\end{equation}
where $\Gamma^\alpha(E)$ is given by Eq.~(\ref{eq:GammaMatrix}) and
\begin{equation}
[C^{\nu  \nu'}]_{n\sigma,m\sigma'}
= \langle \nu | d_{n\sigma} |\nu'\rangle \langle\nu'| d_{m\sigma'}^\dagger |\nu\rangle
\label{eq:Cnm}
\end{equation}
are many-body matrix elements.\cite{Kinaret92,Stafford96c,Stafford98} 
From the normalization of the many-body wavefunctions,
the total resonance width $\tilde{\Gamma}^{\nu\nu'}=\sum_\alpha \tilde{\Gamma}^{\nu  \nu'}_{\alpha}$ scales as
$\mbox{}\sim \sum_\alpha \mbox{Tr}\{\Gamma^\alpha\}/N$, where
$N$ is the number of atomic orbitals in the molecule.
For strongly-correlated systems, there is an additional exponential suppression of Eq.\ (\ref{eq:Cnm}) as $N\rightarrow\infty$
due to the orthogonality catastrophe.\cite{Stafford98,Anderson67}

Linear response transport is determined by the equilibrium Green's functions.
In equilibrium, 
the solution of the set of Eqs.~(\ref{eq:detailed_balance}) reduces to 
\begin{equation}
{\cal P}(\nu) = e^{-\beta(E_\nu - \mu N_\nu)}/{\cal Z},
\label{eq:Pnu_eq}
\end{equation}
where ${\cal Z}$ is the grand partition function of the molecule at inverse temperature $\beta$ and chemical potential $\mu$.

Eq.\ (\ref{eq:Gintra}) implicitly defines the Coulomb self-energy matrix $\Sigma_{\rm C}^{(0)}$ in the sequential tunneling limit via
\begin{equation}
G_{\rm mol}(E) = \left[{\bf 1}E - H_{\rm mol}^{(1)} - \Sigma_{\rm C}^{(0)}(E)\right]^{-1}.
\label{eq:Gmol}
\end{equation}
$\Sigma_{\rm C}^{(0)}$ is the high-temperature 
limit of the Coulomb self-energy
(i.e., the limit $k_B T/\mbox{Tr}\{\Gamma^\alpha\} \gg 1$), and describes intramolecular correlations and charge quantization effects 
(Coulomb blockade away from resonance).
The nonperturbative treatment of intramolecular correlations 
provided by exact diagonalization of $H_{\rm mol}$ in Eq.\ (\ref{eq:Gintra})
allows for an accurate description of the HOMO-LUMO gap, which is essential for a quantitative theory of transport.

\subsection{Correction to the Coulomb self-energy} 
\label{sec:corr_Coulomb}

\begin{figure}[tb]
\centering
\includegraphics[width=3.25in]{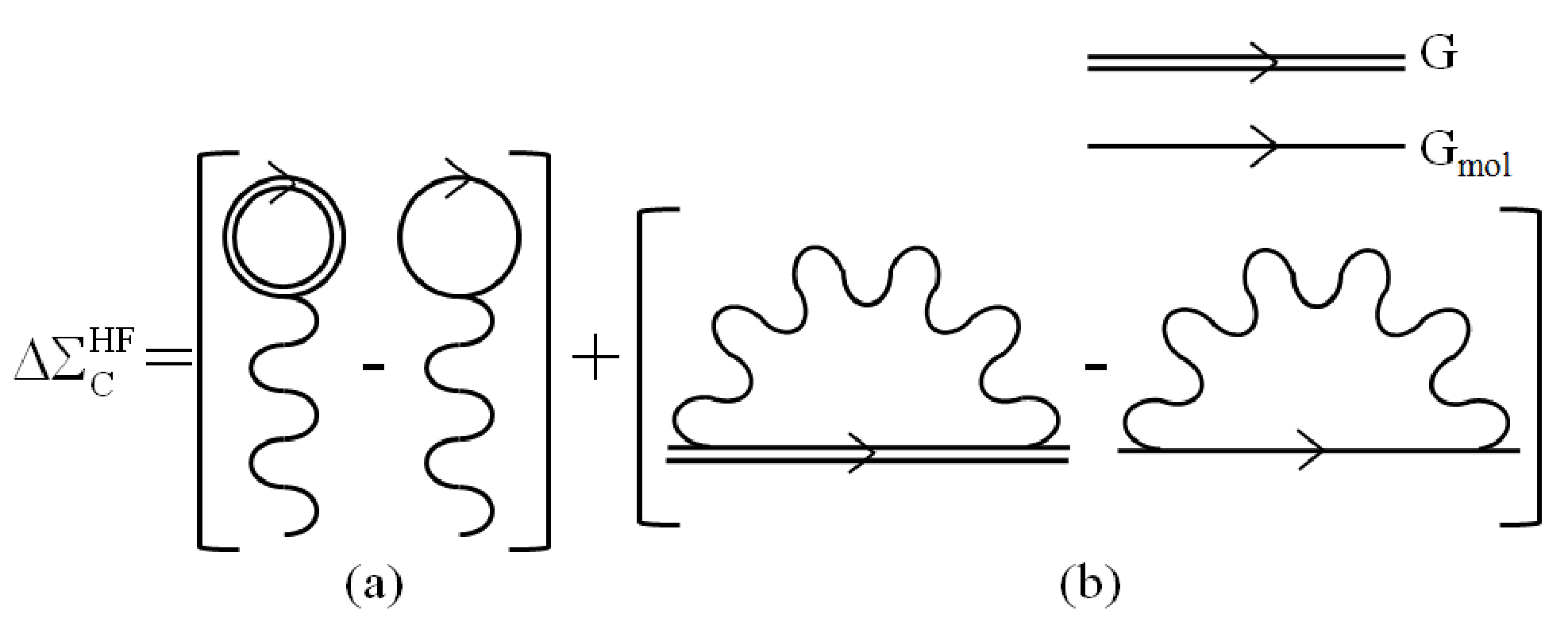} 
\caption[Hartree-Fock self-energy correction]{The correction to the Coulomb self-energy $\Delta\Sigma_{\rm C}$ 
is given in the self-consistent Hartree-Fock approximation by the sum of 
(a) the Hartree (direct) term and (b) the Fock (exchange) term.  The wavy line represents the Coulomb interaction,
which has the form $U_{nm}\delta(\tau-\tau')$ on the Keldysh time contour.}
\label{fig:HF_self_energy}
\end{figure}

In general, the Coulomb self-energy matrix 
$\Sigma_{\rm C} = \Sigma_{\rm C}^{(0)} + \Delta\Sigma_{\rm C}$,
where
$\Delta\Sigma_{\rm C}$ describes the change of the Coulomb
self-energy due to lead-molecule coherence emerging at temperatures $k_B T \lesssim \mbox{Tr}\{\Gamma^\alpha\}$.
Using this decomposition of the Coulomb self-energy, 
Dyson's equation (\ref{eq:Dyson}) can be rewritten in the following useful form:
\begin{equation}
G^{-1}(E) = G_{\rm mol}^{-1}(E) - \Sigma_{\rm T} - \Delta \Sigma_{\rm C},
\label{eq:Dyson2}
\end{equation}
where the self-energy terms $\Sigma_{\rm T}+ \Delta \Sigma_{\rm C}$ describe the effects of finite tunneling width. 
$\Sigma_{\rm T}$ is given by Eq.\ (\ref{eq:SigmaRet_t}). 
Here we point out 
that $\Delta\Sigma_{\rm C}$---unlike $\Sigma_{\rm C}^{(0)}$---can be evaluated 
perturbatively using diagrammatic techniques on the Keldysh time-contour (cf.\ Fig.\ \ref{fig:HF_self_energy}). 
Such a perturbative approach is valid, in principle, 
at temperatures/bias voltages satisfying $\max\{T,eV/k_B\} > T_K$, where $T_K$ is the Kondo temperature\cite{Glazman04}---or 
when there is no unpaired electron on the molecule (such as within the HOMO-LUMO gap of conjugated organic molecules).

$\Sigma_{\rm T}+\Delta\Sigma_{\rm C}$ may be thought of as the response of the junction to turning on the tunneling coupling $\Gamma^\alpha$.
A subtlety in the perturbative evaluation of $\Delta\Sigma_{\rm C}$ is that the diagrams determining the Coulomb self-energy are
typically formulated\cite{Mahan} in terms of the Green's functions of the noninteracting system,
$G^{(0)}(E) = ({\bf 1}E - H_{\rm mol}^{(1)} -\Sigma_{\rm T})^{-1}$
and 
$G^{(0)}_{\rm mol}(E) = ({\bf 1}E - H_{\rm mol}^{(1)} + i 0^+)^{-1}$, while the response of the junction is determined by the full
Green's function $G(E)$ [cf.\ Eq.\ (\ref{eq:Meir_Wingreen})]. 
Evaluating the diagrams in Fig.\ \ref{fig:HF_self_energy} using $G^{(0)}(E)$ and $G^{(0)}_{\rm mol}(E)$ would yield a
correction to the Coulomb self-energy with a pole structure unrelated to that of $\Sigma_{\rm C}^{(0)}$, so that adding the two together
would not yield a physically meaningful result.
Our strategy is thus to calculate $\Delta\Sigma_{\rm C}$ by reformulating the terms in 
the perturbative expansion in terms of the
full Green's function $G(E)$ via appropriate resummations.  This procedure is in general nontrivial, but the result for the Hartree-Fock correction is
given in Sec.\ \ref{sec:HF} below, based on physical arguments. Higher-order self-energy diagrams can be included 
in a similar fashion. 
Electron-phonon coupling\cite{Mitra04,Galperin04,Paulsson05,Viljas05,Vega06,Solomon06} 
can also be included in the many-electron theory via the self-energy terms $\Sigma_{\rm e-ph}
+ \Delta \Sigma_{\rm C}^{\rm e-ph}$, where $\Sigma_{\rm e-ph}$ is given by the usual 
self-energy diagrams,\cite{Galperin04,Paulsson05,Viljas05,Vega06,Solomon06} and $\Delta \Sigma_{\rm C}^{\rm e-ph}$ is the corresponding
correction to the Coulomb self-energy.

\subsubsection{Elastic cotunneling approximation: $\Sigma = \Sigma_{\rm T} + \Sigma_{\rm C}^{(0)}$}
\label{sec:cotunneling}

Far from transmission resonances and for $T\gg T_K$, $\Delta\Sigma_{\rm C}$ can be neglected. 
This is the limit of {\em elastic cotunneling}.\cite{Averin90,Groshev91} 
The Green's functions are given by Eqs.\ (\ref{eq:Dyson}) and (\ref{eq:Keldysh}),
with $\Sigma_{\rm C} = \Sigma_{\rm C}^{(0)}$.
Note that
$[\Sigma_{\rm C}^{(0)}]^<$ does not make a finite contribution to Eq.\ (\ref{eq:Keldysh}) when $\Sigma_{\rm T}$ is finite.
The full NEGF current expression (\ref{eq:Meir_Wingreen}) then reduces to the 
multi-terminal B\"uttiker formula \cite{Buttiker86}
\begin{equation}
\label{eq:Buttiker}
I_\alpha=\frac{2e}{h} \sum_{\beta=1}^M\int_{-\infty}^\infty
dE\;T_{\beta\alpha}(E)\left[f_\beta(E)-f_\alpha(E)\right],
\end{equation}
where the transmission function is given by\cite{Datta95}
\begin{equation}
T_{\beta\alpha}(E)={\rm Tr}\left\{ \Gamma^\alpha(E) G(E) \Gamma^\beta(E) G^\dagger(E)\right\}.
\label{eq:transmission}
\end{equation}
The elastic cotunneling approximation
is  a {\em conserving approximation}---current
is conserved [cf.\ Eq.\ (\ref{eq:Buttiker})] and the spectral function obeys the usual sum-rule. 

\subsubsection{Self-consistent Hartree-Fock correction: $\Sigma_{\rm C} = \Sigma_{\rm C}^{(0)} + \Delta\Sigma_{\rm C}^{\rm HF}$}
\label{sec:HF}

In the 
Hartree-Fock (HF) approximation, the retarded Coulomb self-energy matrix is given by
\begin{equation}
\left[\Sigma_{\rm C}^{\rm HF}\right]_{n\sigma,m\sigma'} = \delta_{\sigma\sigma'} 
\left[\delta_{nm} \sum_{n'} U_{nn'} \langle \rho_{n'} \rangle - U_{nm} \langle d^\dagger_{m\sigma} d_{n\sigma} \rangle \right],
\label{eq:Sigma_HF}
\end{equation}
and $\left[\Sigma_{\rm C}^{\rm HF}\right]^< = 0$.
In general,
\begin{equation}
\langle d_{n\sigma}^\dagger d_{m\sigma'}\rangle = -\frac{i}{2\pi} \int_{-\infty}^\infty dE \, [G^<(E)]_{m\sigma', n\sigma}
\label{eq:corr_gen}
\end{equation}
and
\begin{equation}
\lim_{\Sigma_{\rm T} \rightarrow 0} \langle d_{n\sigma}^\dagger d_{m\sigma'}\rangle 
= 
\sum_{\nu} {\cal P}(\nu) \langle \nu | d_{n\sigma}^\dagger d_{m\sigma'} | \nu\rangle.
\label{eq:corr_mol}
\end{equation}
The Hartree-Fock correction is then
$\Delta\Sigma_{\rm C}^{\rm HF} \equiv \Sigma_{\rm C}^{\rm HF} - 
\Sigma_{\rm C}^{\rm HF}|_{\Sigma_{\rm T} \rightarrow 0} $.
The Feynman diagrams representing this correction are shown in Fig.~\ref{fig:HF_self_energy}.  
A self-consistent solution of Eqs.\ 
(\ref{eq:Keldysh}), (\ref{eq:Dyson2}), and (\ref{eq:Sigma_HF})--(\ref{eq:corr_mol}) yields
a conserving approximation in which the junction current is again given by Eq.\ (\ref{eq:Buttiker}).

\begin{figure}[tb] 
    \includegraphics[height=8.0cm,angle=-90]{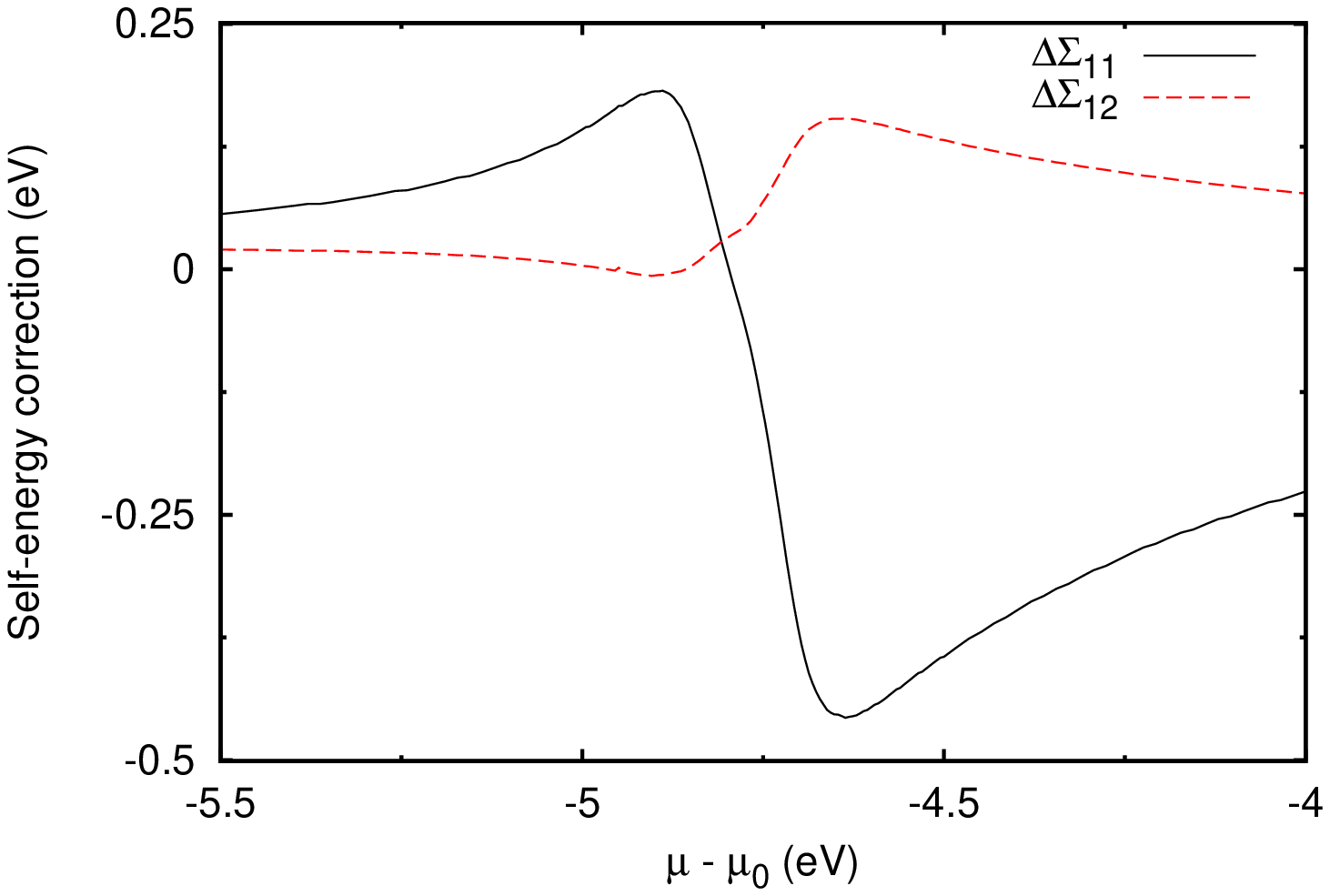}
   \caption{Self-consistent Hartree-Fock correction to the Coulomb self-energy matrix of a diatomic molecule versus lead chemical potential, 
shown in the vicinity of the HOMO resonance.  Here 
$\mu_0= \left(\varepsilon_{\rm HOMO}+\varepsilon_{\rm LUMO}\right)/2$ is the chemical potential of the isolated molecule,
$T=300\mbox{K}$, and $\Gamma_1=\Gamma_2=0.2\mbox{eV}$
(see Sec.\ \ref{sec:molecular_model}).
  }
  \label{fig:Sigma_HF}
\end{figure}

The use of the interacting Green's functions $G$ and $G_{\rm mol}$ 
in the evaluation of the Hartree self-energy is clearly justified on physical grounds, since this yields the classical 
electrostatic potential due to the actual nonequilibrium charge distribution on the molecule.  
The direct (Hartree) and exchange (Fock) contributions to $\Delta \Sigma_{\rm C}$ must be treated on an equal footing in order
to cancel the unphysical self-interaction, justifying the use of the same interacting Green's functions in the evaluation of
the exchange self-energy.
The diagrams of Fig.~\ref{fig:HF_self_energy} would be quite complex if expressed in terms of noninteracting Green's functions,
because both $G$ and $G_{\rm mol}$ involve $\Sigma_{\rm C}^{(0)}$, which includes all possible combinations of Coulomb lines and
intramolecular propagators.

\begin{figure}[bt]
\centering
\includegraphics[angle=-90,width=3.25in]{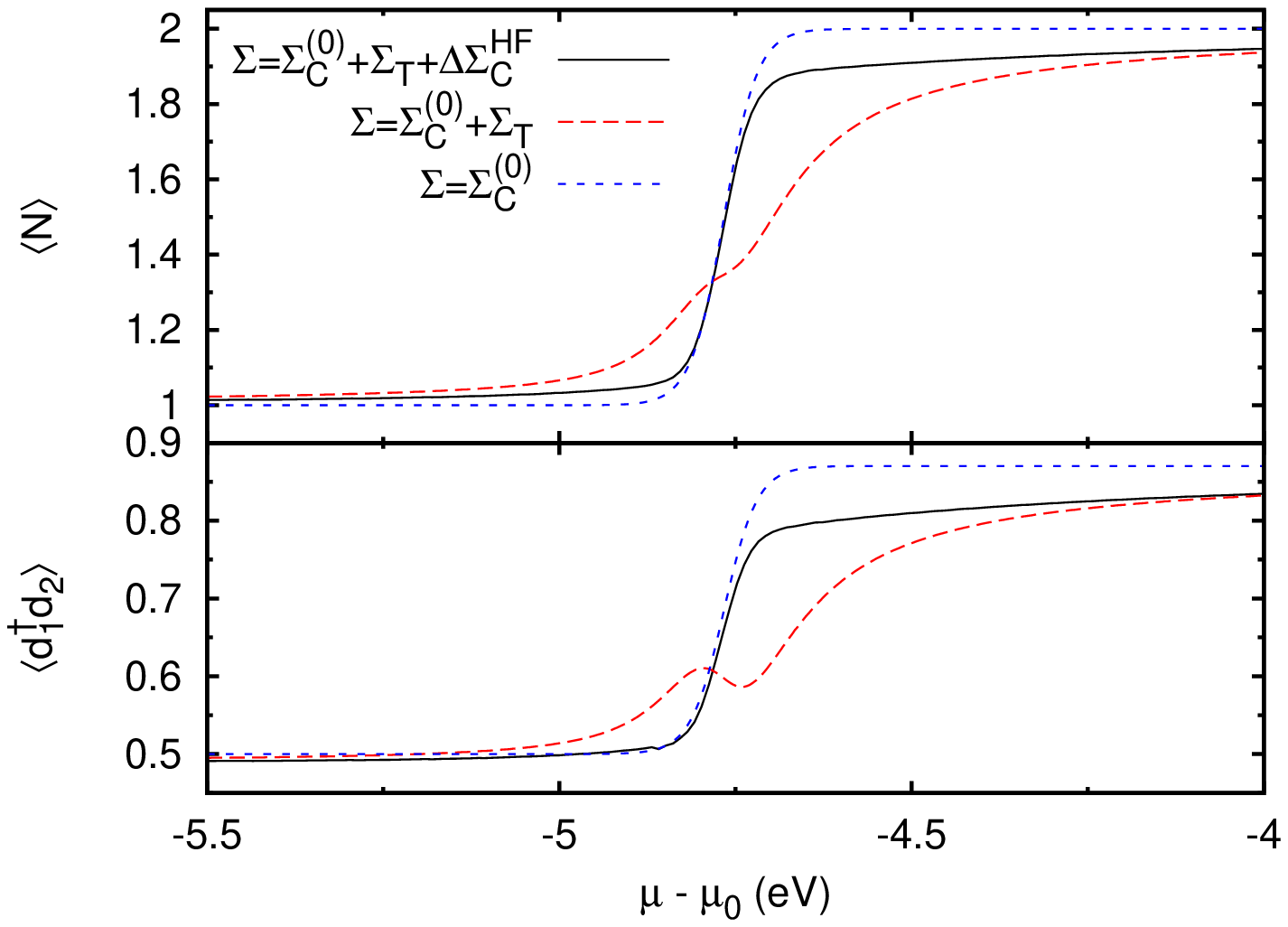} 
\caption{Equilibrium correlation functions near the HOMO resonance for a diatomic molecule 
(same parameters as in Fig.\ \ref{fig:Sigma_HF}).  
Top:  Total molecular charge $\langle N \rangle$ in three different approximations: $\Sigma=\Sigma_{\rm C}^{(0)}$ (isolated
molecule in the grand canonical ensemble); $\Sigma=\Sigma_{\rm C}^{(0)}+\Sigma_{\rm T}$ (elastic cotunneling approximation); and
$\Sigma=\Sigma_{\rm C}^{(0)}+\Sigma_{\rm T}+\Delta\Sigma_{\rm C}^{\rm HF}$ (self-consistent HF 
correction).
The charging step is broadened in the elastic cotunneling approximation, compared to that of the isolated molecule, and acquires a
`knee' near resonance.  In the self-consistent HF approximation, the charge imbalance 
$\Delta N = \langle N \rangle - \langle N \rangle|_{\Sigma_{\rm T}\rightarrow 0}$ is strongly screened near resonance, leading
to a charging step as steep as that of the isolated molecule, but asymptotically approaches the elastic cotunneling result away from resonance.
Bottom:  The correlation function $\langle d^\dagger_{1\sigma}d_{2\sigma}\rangle$ 
in the same three approximations.}
\label{fig:diatomic_nandcorr_HF}
\end{figure}

The self-consistent Hartree-Fock correction for a diatomic molecule is
shown in Fig.\ \ref{fig:Sigma_HF}. 
The parameters were chosen so that the resonance width $\mbox{}\sim \sum_\alpha \mbox{Tr}\{\Gamma^\alpha\}/N=0.2\mbox{eV}$, where
$N$ is the number of atomic orbitals in the molecule.
The elements of the matrix $\Delta \Sigma_{\rm C}^{\rm HF}$ are
largest near a transmission resonance, but vanish on resonance.
This behavior can be understood by considering the molecular correlation functions shown in Fig.\ \ref{fig:diatomic_nandcorr_HF}.
Inclusion of the tunneling self-energy without the corresponding correction to the Coulomb self-energy leads to a charge imbalance on the molecule
near resonance.  This in turn leads to a Hartree correction which tends to counteract the charge imbalance. A corresponding behavior 
is found for the exchange correction and off-diagonal correlation function.
A non-self-consistent calculation would yield a 
much larger correction $\Delta \Sigma_{\rm C}^{\rm HF}$,
indicating the important role of screening.  
It should be pointed out that a treatment of screening in linear response is not adequate near resonance.

\begin{figure}[b] 
    \includegraphics[height=8.0cm,angle=-90]{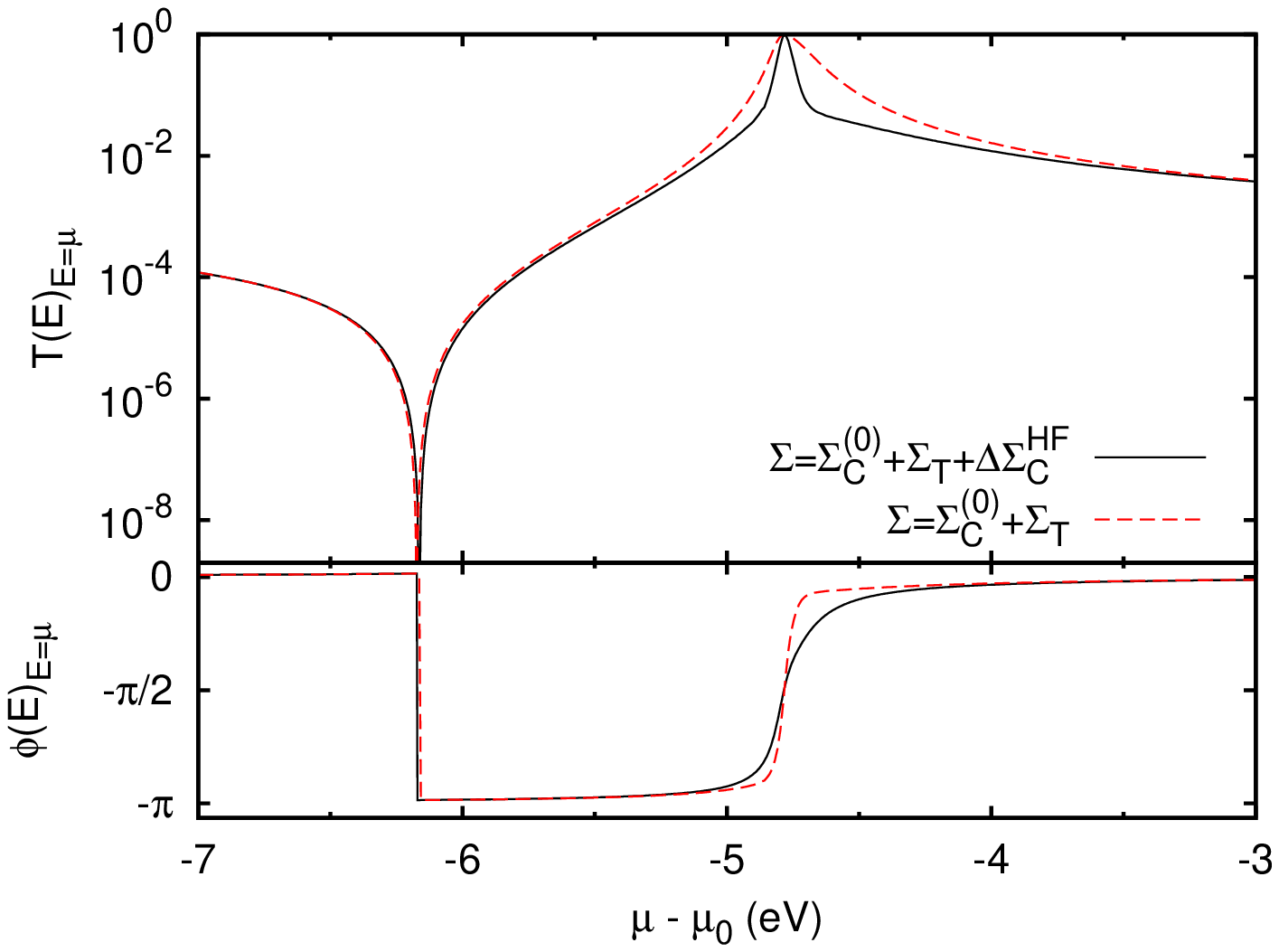}
   \caption{
Transmission probability and phase 
calculated with and without 
the self-consistent HF 
correction. 
Note that the transmission peaks and nodes are not shifted by $\Delta\Sigma_{\rm C}^{\rm HF}$, 
but the width of the transmission resonance is reduced asymmetrically.
  }
  \label{fig:T_and_phase_HF}
\end{figure}

As shown in Fig.\ \ref{fig:T_and_phase_HF},
the transmission peaks and nodes are not shifted by the self-consistent HF correction, and the transmission phase is not changed significantly.
Transport properties [cf.\ Eqs.\ (\ref{eq:Buttiker}) and (\ref{eq:transmission})]
are therefore qualitatively better described in the elastic cotunneling approximation than are the correlation functions, and the approximation
is quantitatively accurate in the cotunneling regime $|\mu - \mu_{\rm res}| > \Gamma$, $\max\{T,eV/k_B\} > T_K$.

The tendency toward charge quantization near resonance is significantly
increased by the self-consistent HF correction, as shown in Fig.\ \ref{fig:diatomic_nandcorr_HF}. 
The steepness of the self-consistent charging step is limited only by thermal broadening.  This result is consistent with previous
theoretical studies\cite{Golubev94,Matveev95,Goppert01} of Coulomb blockade in metal islands and quantum dots, but inconsistent with
the behavior of the Anderson model,\cite{Glazman04} where singular spin fluctuations modify this generic behavior.

\section{Molecular Heterojunction Model}
\label{sec:molecular_model}

Heterojunctions formed from 
$\pi$-conjugated molecules are of particular interest, both because of their relevance for nanotechnology\cite{Tao06,Cardamone06} 
and their extensive experimental characterization.\cite{Nitzan03,Natelson06,Tao06} 
A semi-empirical $\pi$-electron Hamiltonian \cite{Chandross97,Castleton02,Cardamone06}
can be used to model the electronic degrees of
freedom most relevant for transport:
\begin{eqnarray}
\label{eq:H_mol}
 H_{\rm mol}=\sum_{n,\sigma}\varepsilon_n d_{n\sigma}^\dagger d_{n\sigma}
-\sum_{n,m,\sigma}\left(t_{nm}d_{n\sigma}^\dagger
  d_{m\sigma}+\mathrm{H.c.}\right)
\mbox{ }+ 
\sum_{n,m}\frac{U_{nm}}{2}Q_nQ_m,\,\,\,\,
\end{eqnarray}
where $d^\dagger_{n\sigma}$ creates an electron of spin $\sigma$ in the $\pi$-orbital of the $n$th carbon atom, $\varepsilon_n$ is the atomic orbital 
energy, and $t_{nm}$ is the hopping matrix element between orbitals $n$ and $m$.  
In the $\pi$-electron theory, the effect of different side-groups is included through shifts of the orbital energies $\varepsilon_n$. 
The effect of substituents (e.g., thiol groups) used to bond the leads to the molecule
can be included\cite{Tian98,Nitzan01} in the tunneling matrix elements $V_{nk}$ [cf.\ Eq.\ (\ref{eq:Htun})]. 

The effective charge operator for orbital $n$ is\cite{Stafford98,Cardamone06} 
\begin{equation}
Q_n=\sum_\sigma d_{n\sigma}^\dagger d_{n\sigma}-\sum_\alpha C_{n\alpha}\mathcal{V}_\alpha/e -1,
\label{eq:Qn}
\end{equation}
where
$C_{n\alpha}$ is the capacitive coupling between orbital $n$ and lead $\alpha$, $e$ is the electron charge, and $\mathcal{V}_\alpha$ is the voltage on 
lead $\alpha$.  The lead-molecule capacitances are elements of a full capacitance matrix 
$C$, 
which also includes the intra-molecular 
capacitances, determined by the relation $C_{nm}=e^2[U^{-1}]_{nm}$.  
The values $C_{n\alpha}$ are determined by the zero-sum rules required for gauge invariance\cite{CapMatrixBook84} 
\begin{equation}
\sum_m C_{nm} + \sum_{\alpha=1}^M C_{n\alpha}=0,
\label{eq:zero_sum_rule}
\end{equation}
and by the geometry of the junction (e.g.\ $C_{n\alpha}$ inversely proportional to lead-orbital distance).

The effective interaction energies for $\pi$-conjugated systems can be written\cite{Chandross97,Castleton02}
\begin{equation}\label{eq:ohno}
U_{nm}=\delta_{nm}U_0+\left(1-\delta_{nm}\right)\frac{U_0}{\epsilon  \sqrt{1+\alpha(R_{nm}/\mbox{\AA})^2}},
\end{equation}
where $U_0$ is the on-site Coulomb repulsion, $\alpha=\left( U_0/14.397\mbox{eV} \right)^2$,
and $R_{nm}$ is the distance between orbitals $n$ and $m$.  The phenomenological 
dielectric constant $\epsilon$ accounts for screening due to both the 
$\sigma$-electrons and any environmental considerations, such as non-evaporated solvent.\cite{Castleton02}  
With an appropriate choice of the parameters $t_{nm}$, $U_0$, and $\epsilon$, the {\em complete spectrum of electronic excitations} up to
8--10eV of the molecules
benzene, biphenyl, and {\em trans-}stilbene in the gas phase can be 
reproduced with high accuracy\cite{Castleton02}
by exact diagonalization of Eq.\ (\ref{eq:H_mol}).  
An accurate description of excited states is essential to model transport far from equilibrium. 
Larger conjugated organic molecules can also be 
modeled \cite{Chandross97,Chakrabarti99} via Eqs.\ (\ref{eq:H_mol}) and (\ref{eq:ohno}).

$\sigma$-orbitals can also be included in Eq.\ (\ref{eq:H_mol}) as additional energy bands, and the
resulting multi-band extended Hubbard model can be treated using the same NEGF formalism sketched above.  
Tunneling through the $\sigma$-orbitals may be important in small molecules,\cite{Ke08,Solomon08a}
especially in cases where quantum interference leads to a transmission node in the $\pi$-electron system. 

The biggest uncertainty in modelling single-molecule heterojunctions is the lead-molecule coupling.\cite{Baranger05}
For this reason, we take the two most uncertain quantitites characterizing lead-molecule coupling---the tunneling width $\Gamma$ and the
chemical potential offset $\Delta \mu$ of isolated molecule and metal electrodes---as phenomenological parameters to be determined by fitting
to experiment.  In the broad-band limit\cite{Jauho94} for the metallic electrodes, and assuming each electrode is covalently bonded to
a single carbon atom of the molecule, the tunneling-width matrix reduces to a single constant:
$\Gamma_{nm}^\alpha(E)=\Gamma_\alpha \delta_{na}\delta_{ma}$, where $a$ is
the orbital connected to electrode $\alpha$.  
Typical estimates \cite{Mujica94,Tian98,Nitzan01} 
indicate $\Gamma \lesssim 1\mbox{eV}$ for organic molecules coupled to gold contacts via thiol groups.

\section{Benzene(1,4)dithiol junction}
\label{sec:BDT}

As a first application of our many-body theory of molecular junction transport,
we consider the benchmark system of 
benzene(1,4)dithiol (BDT) with two gold 
leads.\cite{Reed97,Reichert03,Xiao04,Dadosh05,Kriplani06,Toher07,Reddy07,Baheti08}  
The Hamiltonian parameters for benzene are\cite{Castleton02} 
$U_0=8.9\mbox{eV}$, $\epsilon=1.28$, 
$t_{nm}=2.68\mbox{eV}$ for $n$ and $m$ nearest neighbors, and $t_{nm}=0$ otherwise.  
We consider a symmetric junction (symmetric capacitive couplings and $\Gamma_1=\Gamma_2\equiv \Gamma$)
at room temperature (T=300K).

\subsection{Linear electric and thermoelectric junction response}
\label{sec:linresp}

Thermoelectric effects \cite{Paulsson03b,Reddy07,Baheti08} provide important insight into the transport mechanism in single-molecule
junctions,  but are particularly sensitive to correlations,\cite{Chaikin76,Stafford93b}
calling into question the applicability of single-particle or mean-field theory.
We can now investigate thermoelectric effects in molecular heterojunctions for the first time using many-body theory.
The thermopower $S$ of a molecular junction is obtained by measuring the voltage $\Delta {\cal V}\equiv -S\Delta T$ created across an open junction in 
response to a temperature differential $\Delta T$.  
In general, $S$ can be calculated by taking the appropriate linear-response limit of
Eq.\ (\ref{eq:Meir_Wingreen}), which includes both elastic and inelastic processes.
However, for purely elastic transport [cf.\ Secs.\ \ref{sec:cotunneling} and \ref{sec:HF}], Eq.\ (\ref{eq:Buttiker}) can be used to derive
the well-known result\cite{Sivan86,vanHouten92}
\begin{equation}
\label{eq:thermopower}
S(\mu,T)=-\frac{1}{\mbox{eT}}\frac{ \int_{-\infty}^\infty  T_{12}(E) \left( -\frac{\partial f}{\partial E} \right) 
\left( E-\mu \right) dE}{\int_{-\infty}^\infty  T_{12}(E) \left( -\frac{\partial f}{\partial E} \right)dE }.
\end{equation}

\begin{figure}[tb]
\centering
\includegraphics[width=3.2in]{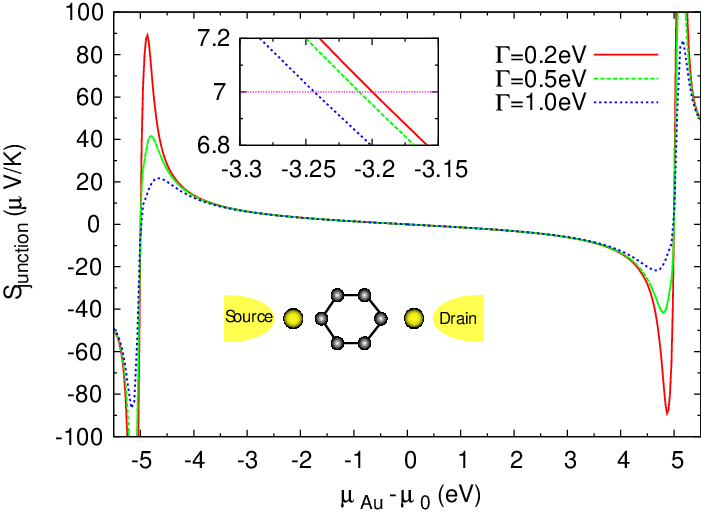} 
\caption[BDT Thermopower]{
Thermoelectric power of a BDT-Au junction at $T=300$K as a function of the lead-molecule chemical potential mismatch for three different
tunneling widths. 
Comparison to the experimental value\cite{Baheti08} $7.0\pm0.2\mu \mbox{V}/\mbox{K}$ 
fixes $\mu_{\rm Au}=\mu_0 -(3.22 \pm 0.04)\mbox{eV}$ (see inset showing
closeup of experimentally relevant parameter range).
}
\label{fig:thermopower}
\end{figure}

Thermopower 
measurements\cite{Baheti08,Reddy07} provide a 
means to determine\cite{Paulsson03b} the lead-molecule chemical potential mismatch $\Delta \mu=\mu_{\rm Au}-\mu_0$,  
where $\mu_0= \left(\varepsilon_{\rm HOMO}+\varepsilon_{\rm LUMO}\right)/2$ and $\varepsilon_{\rm HOMO}$ ($\varepsilon_{\rm LUMO}$) is the HOMO (LUMO) 
energy level.  
Figure \ref{fig:thermopower} shows the thermopower of a BDT junction as a function of
$\Delta \mu$, calculated from Eqs.\ (\ref{eq:transmission}) and
(\ref{eq:thermopower}) in the elastic cotunneling approximation.
As pointed out by Paulsson and Datta,\cite{Paulsson03b} 
$S$ is nearly independent of $\Gamma$ away from the transmission resonances, 
allowing $\Delta \mu$ to be determined directly by comparison to experiment. 
The thermopower of a BDT-Au junction was recently measured by Baheti {\it et al.,}\cite{Baheti08} who obtained the result
$S=(7.0\pm0.2)\mu \mbox{V}/\mbox{K}$.
Equating this experimental value with the calculated thermopower shown in Fig.\ \ref{fig:thermopower},
we find that $-3.25{\rm eV}\leq \mu_{\rm Au}-\mu_0\leq-3.15{\rm eV}$ over a broad range of $\Gamma$ values.  
The Fermi level of gold thus lies about $1.8\mbox{eV}$ above the HOMO resonance, validating the notion that 
transport in these junctions is hole-dominated.
Nonetheless, $\mu_{\rm Au}-\varepsilon_{\rm HOMO}$ is sufficiently large that the elastic cotunneling approximation is well justified.

The only other free parameter in the molecular junction model is the tunneling width $\Gamma$, which can be found by matching 
the linear-response conductance to experiment.  
Although there is a large range of experimental values,\cite{Lindsay07} the most reproducible and lowest resistance contacts were obtained by 
Xiao, Xu and Tao,\cite{Xiao04} who report a single-molecule conductance value of 0.011G$_0$.  
This fixes $\Gamma=(0.63\pm0.02)\mbox{eV}$ with $\mu_{\rm Au}-\mu_0=-3.22\pm0.04\mbox{eV}$. This value of $\Gamma$ is
within the range predicted by other groups for similar molecules.\cite{Mujica94,Tian98,Nitzan01,Baranger05}
With the final parameter in the model 
fixed, we can now use our many-body transport theory to predict the linear and nonlinear response of this molecular junction.

\begin{figure*}[htb]
\centering
\includegraphics[width=7in]{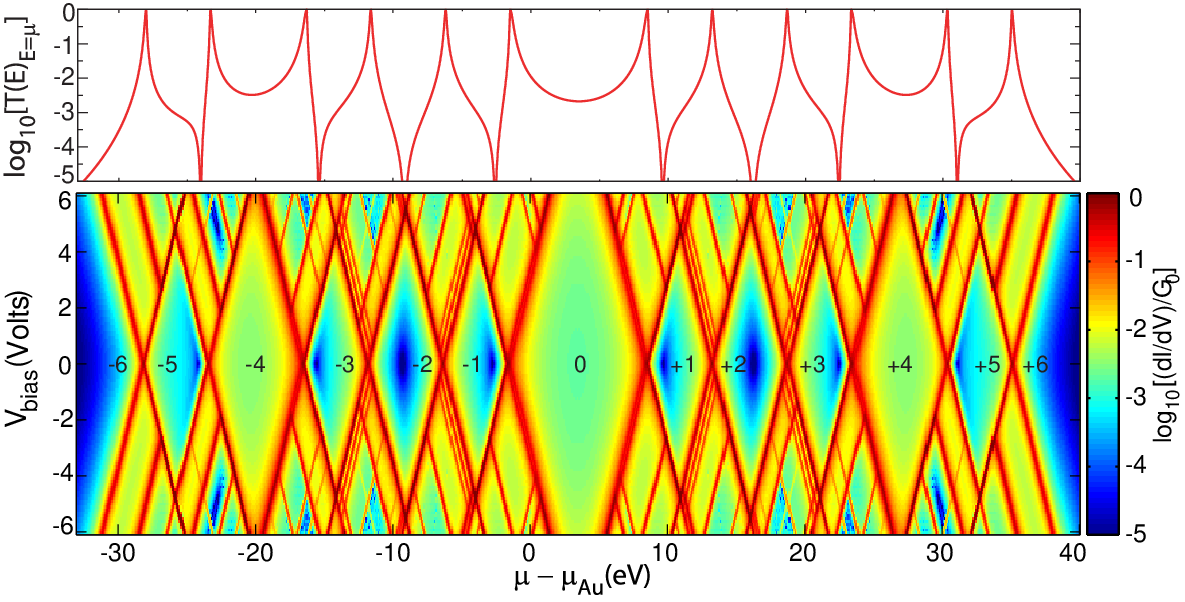}
\caption[Molecular Diamond]{
Linear-response conductance versus chemical potential (top panel) and
differential conductance 
versus chemical potential and bias voltage (bottom panel)
for a 
benzene(1,4)dithiol--Au 
junction. The calculation was carried out within the elastic cotunneling approximation using Eq.\ (\ref{eq:Buttiker}), 
with $\Gamma_1 = \Gamma_2 = 0.6\mbox{eV}$ and $T = 300\mbox{K}$.
Here $G_0=2e^2/h$ is the conductance quantum.
Note the asymmetric Fano-like lineshapes 
and the transmission nodes (arising from destructive interference) in the top panel.
The numbers within the central diamonds of the bottom panel indicate the quantized  charge on the molecule (relative to neutral benzene).
The full spectrum is shown for completeness.}
\label{fig:FullCoulombDiamond}
\end{figure*}

The transmission probability $T(E)|_{E=\mu}$ of the BDT junction is shown as a function of $\mu-\mu_{\rm Au}$ 
in the upper panel of Fig.\ \ref{fig:FullCoulombDiamond}.
This is the linear-response conductance in units of the conductance quantum $G_0=2e^2/h$.
The chemical 
potential $\mu$ is related to the gate voltage $V_g$ in a three-terminal junction via the gate capacitance $C_g$.  
The transmission 
spectrum shown in 
Fig.\ \ref{fig:FullCoulombDiamond} exhibits several striking features: 
a large but irregular peak spacing with an increased HOMO-LUMO gap, 
non-symmetric, Fano-like resonance lineshapes, and transmission nodes due to destructive interference in the coherent quantum transport.

\subsection{Non-linear junction transport}
\label{sec:nonlin}

We next calculate the differential conductance $\partial I / \partial V_{\rm bias}$ of the BDT-Au junction
as a function of $\mu$ and $V_{\rm bias}$
(see Fig.\ \ref{fig:FullCoulombDiamond}, lower panel).  
The current was calculated in the elastic cotunneling approximation 
using Eqs.\ (\ref{eq:Buttiker}) and (\ref{eq:transmission}).
This approximation accurately
describes nonresonant transport, including the transmission nodes, as well as the positions and heights of the transmission resonances,
as discussed in Sec.\ \ref{sec:HF}.
The differential conductance spectrum 
of the junction exhibits
clear signatures of excited-state transport \cite{Weis93,DeFranceschi01} and an irregular ``molecular diamond'' structure analogous to the
regular Coulomb diamonds observed in quantum dot transport experiments.\cite{DeFranceschi01}
The charge on the molecule is quantized within the central diamonds of Fig.\ \ref{fig:FullCoulombDiamond},
an important interaction effect inaccessible to so-called {\em ab initio} mean-field calculations.
In Fig.\ \ref{fig:FullCoulombDiamond}, the full spectrum is shown for completeness, although the junction may not be stable over the
entire range of bias and gate voltages shown.

Apart from the central HOMO-LUMO gap, the widths of the diamonds
in Fig.~\ref{fig:FullCoulombDiamond} 
can be roughly explained via a capacitive model in which the molecule 
is characterized by a single capacitance $C_{\rm mol}=e^2/\left\langle U_{nm}\right\rangle$, where 
$\left\langle U_{nm}\right\rangle = 5.11\mbox{eV}$ is the
average over all molecular sites.  
The HOMO-LUMO gap of $\sim$10eV is significantly larger than this estimate of the charging energy,
an indication of the significant deviations of our theory from a simple constant interaction 
model.  

Charge quantization, also known as {\em Coulomb blockade}, has 
been observed in several different types of molecular heterojunctions,\cite{Park00,Liang02,Kubatkin03,Poot06,Danilov08} but has
not yet been observed in BDT junctions due to the difficulty of gating such small molecules.\cite{Li06} 
The unambiguous observation of Coulomb blockade in 
junctions involving larger molecules
(with {\em smaller} charging energies) indicates that 
such interaction effects, lying outside the scope of mean-field approaches, are undoubtedly even more pronounced in small molecules like BDT. 
Important aspects of this phenomenon remain to be understood in larger molecules, such as the anomalously low
reported values of the charging energy.\cite{Park00,Liang02,Kubatkin03,Poot06,Danilov08}

A zero-bias cross section of the bottom panel of Fig.~\ref{fig:FullCoulombDiamond} reproduces the transmission 
spectrum shown in 
the top panel of the same figure.  Increasing the bias voltage, we find that the resonances split into negatively sloped particle-like 
($|e|dV/d\mu = -\left(C_1+C_2+C_g\right)/C_1$) lines and positively sloped hole-like ($|e|dV/d\mu = +\left(C_1+C_2+C_g\right)/C_2$) lines, 
where $C_1$ and $C_2$ are the mean lead-molecule capacitances, defined by
$C_\alpha= \langle C_{n\alpha}\rangle$, $\alpha=1,2$.  In the symmetric coupling case ($C_1=C_2$ with $C_g/C_1\ll 1$) the lines 
therefore have slopes of -2 and +2 for particle-like and hole-like lines, respectively.  Within the V-shaped outline traced by the particle-like 
and hole-like lines, we find signatures of resonant tunneling through electronic excited states in the many narrow, nearly parallel resonance lines.  
While transport through electronic excited states has not yet been unambiguously identified in single-molecule heterojunctions, it has been 
observed in 
quantum dots\cite{Weis93,DeFranceschi01} and carbon nanotubes.\cite{Sapmaz06} 

An accurate description of the HOMO-LUMO gap is essential for a quantitative theory of transport in molecular heterojunctions.
The central HOMO-LUMO gap shown 
in Fig.\ \ref{fig:FullCoulombDiamond} is significantly larger than that predicted\cite{Nitzan03} by density functional
theory---which neglects charge-quantization effects---but is consistent with previous many-body calculations
in the sequential tunneling regime.\cite{Schoeller03,Datta06,Begemann08} 
It should be emphasized that the transport gap $\Delta\mu_{\rm tr}$
in a molecular junction exceeds the optical gap 
$\hbar \omega_{\rm min}$ of an isolated molecule. 
Roughly speaking, $\Delta\mu_{\rm tr}\simeq \hbar \omega_{\rm min}+e^2/C_{\rm mol} + E_x$, where 
$e^2/C_{\rm mol}$ is the charging energy (see discussion above) and $-E_x$ is the exciton binding energy.
Excitonic states of the BDT-Au junction can be identified in the differential conductance spectrum of Fig.\ \ref{fig:FullCoulombDiamond}
as the lowest-energy (i.e., smallest bias) excitations outside the central diamond of the HOMO-LUMO gap, from which it is apparent that
$\Delta\mu_{\rm tr}\approx 2 \hbar \omega_{\rm min}$ for a BDT-Au junction.

\section{Conclusions}
\label{sec:conclusion}

In conclusion, we have developed a many-body theory of electron transport 
in single-molecule heterojunctions 
that treats coherent quantum effects and Coulomb interactions on an equal footing.
As a first application of our theory, we have investigated the thermoelectric power and differential conductance 
of a prototypical single-molecule junction, benzenedithiol with gold electrodes.

Our results reproduce the key features of both the coherent 
and Coulomb blockade transport regimes: 
Quantum interference effects, such as the transmission nodes 
predicted within mean-field theory,\cite{Cardamone06,Solomon08a,Solomon08b}
are confirmed, while 
the differential conductance spectrum exhibits characteristic charge quantization 
``diamonds''\cite{Park00,Liang02,Kubatkin03,Poot06,Danilov08}---an effect 
outside the scope of mean-field approaches based on density-functional theory.
The HOMO-LUMO transport gap obtained is consistent
with previous many-body treatments in the sequential tunneling limit.\cite{Schoeller03,Datta06,Begemann08}

The central object of the many-body theory is the Coulomb self-energy $\Sigma_{\rm C}$ of the junction,
which may be expressed as $\Sigma_{\rm C}=\Sigma_{\rm C}^{(0)}+\Delta\Sigma_{\rm C}$, where $\Sigma_{\rm C}^{(0)}$ is the 
result 
in the sequential tunneling limit, and $\Delta\Sigma_{\rm C}$ is the correction due to a finite tunneling width $\Gamma$. 
In this article, we have evaluated 
$\Sigma_{\rm C}^{(0)}$ exactly, thereby including intramolecular correlations at a nonperturbative level, while
the direct and exchange contributions to 
$\Delta\Sigma_{\rm C}$ were evaluated self-consistently using a conserving approximation based on diagrammatic perturbation theory on the Keldysh
contour.  An important feature of our theory is that this approximation for 
$\Delta\Sigma_{\rm C}$ can be systematically improved by including additional processes diagrammatically.
In this way, important effects such as dynamical screening, spin-flip scattering,\cite{Glazman04} and
electron-phonon coupling\cite{Mitra04,Galperin04,Paulsson05,Viljas05,Vega06,Solomon06}
can be included as natural extensions of the theory.

Our general theory of molecular junction transport should be contrasted to results for transport in the 
Anderson model.\cite{Glazman04,Galperin07} 
The Anderson model provides important insight into nonperturbative interaction effects in quantum transport through nanostructures;
however, the internal structure of the molecule is neglected.
Moreover, since it is limited to a single spin-degenerate level, it can only describe
a single Coulomb diamond with an odd number of electrons on the molecule, and is therefore not applicable to transport within the HOMO-LUMO
gap of conjugated organic molecules, where the number of electrons is even (more precisely, the HOMO-LUMO gap is taken to be infinite
in the Anderson model).

Recently, a NEGF approach for molecular junction transport was introduced
\cite{Galperin08}
using Green's functions that are matrices in a basis of many-body molecular eigenstate outer products $|\nu\rangle\langle \nu'|$;  it is unclear
what relation this approach has to 
the approach proposed herein involving {\em single-particle Green's functions}, which are matrices in the atomic orbital basis.
     
\acknowledgments
The authors thank S.\ Mazumdar and D.\ M.\ Cardamone for useful discussions.


\end{document}